\documentclass[twocolumn,prl,aps,superscriptaddress]{revtex4}
\usepackage{color}
\usepackage{amsmath}
\usepackage{amssymb}
\usepackage{graphicx}
\usepackage{esint}
\usepackage[unicode=true,
 bookmarks=false,
 breaklinks=true,pdfborder={0 0 0},backref=false,colorlinks=true]
 {hyperref}

\makeatletter

\providecommand{\tabularnewline}{\\}

\@ifundefined{textcolor}{}
{
 \definecolor{BLACK}{gray}{0}
 \definecolor{WHITE}{gray}{1}
 \definecolor{RED}{rgb}{1,0,0}
 \definecolor{GREEN}{rgb}{0,1,0}
 \definecolor{BLUE}{rgb}{0,0,1}
 \definecolor{CYAN}{cmyk}{1,0,0,0}
 \definecolor{MAGENTA}{cmyk}{0,1,0,0}
 \definecolor{YELLOW}{cmyk}{0,0,1,0}
}

\usepackage{color}
\usepackage{mathrsfs}
\usepackage{dcolumn}
\usepackage{bm}
\usepackage{amsfonts}

\makeatother

\begin{document}

\title{Valley polarization generated in 3-dimensional group-IV monochalcogenids}

\author{Tiantian Zhang$^{*}$}
\affiliation{Institute of Physics, Chinese Academy of Sciences/Beijing National Laboratory for Condensed Matter Physics, Beijing 100190, China}
\affiliation{Wuhan National High Magnetic Field Center and School of Physics, Huazhong University of Science and Technology, Wuhan 430074, China}
\affiliation{University of Chinese Academy of Sciences, Beijing 100049, China}

\author{Sijie Zhang\footnote{These two authors contributed equally to this work}}
\affiliation{International Center for Quantum Materials, School of Physics, Peking University, China}

\author{Gang Xu\footnote{gangxu@hust.edu.cn}}
\affiliation{Wuhan National High Magnetic Field Center and School of Physics, Huazhong University of Science and Technology, Wuhan 430074, China}

\author{Zhong Fang}
\affiliation{Beijing National Laboratory for Condensed Matter Physics and Institute
of Physics, Chinese Academy of Sciences, Beijing 100190, China}
\affiliation{Collaborative Innovation Center of Quantum Matter, Beijing, 100084, China}

\author{Nanlin Wang}
\affiliation{International Center for Quantum Materials, School of Physics, Peking University, China}
\affiliation{Collaborative Innovation Center of Quantum Matter, Beijing, 100084, China}

\date{\today}
\begin{abstract}
Valleytronics is one of the breaking-through to the technology of electronics, which provides a new degree of freedom to manipulate the properties of electrons. Combining DFT calculations, optical absorption analysis and the linear polarization-resolved transmission measurement together, we report that three pairs of valleys, which feature opposite optical absorption, existing in the 3-dimensional (3D) group-IV monochalcogenids. By applying the linearly-polarized light, valley polarization is successfully generated for the first time in a 3D system, which opens a new direction for the exploration of the valley materials and provides a good platform for the photodetector and valleytronic devices. Valley modulation versus the in-plane strain in GeSe is also studied, suggesting an effective way to get the optimized valleytronic properties.
\end{abstract}

\maketitle
\section{\label{sec:level1}Introduction}

As a consequence of the crystal symmetry, energy spectrum in the semiconductors
may contains many energy-adjacent minima (maxima) in the conduction
(valence) bands, forming multiple valleys that are located along different
axises with different momenta. The electrons trapped in different valleys
usually carry different orbital compositions and momenta \cite{valleypolarization_1,valleyproperty_1,Valleytronics_1,Valleytronics_2},
 meaning the flow of charge in a particular way and leading to
disparate Zeeman splitting and optical absorption \cite{Opticalabsorption_1,Opticalabsorption_2,Opticalabsorption_3,Opticalabsorption_4,Opticalabsorption_5}.
Thus, controlling electrons in the specific valley, $e.g.$, valley
polarization, provides an unprecedented way of tuning optical and
electronic properties in semiconductors \cite{tuningproperties_1,tuningproperties_2},
and creates a new frontier in condensed matter physics, the valleytronics
\cite{Valleytronics_1,Valleytronics_2}, which aims to generate/detect the
valley polarization and constructs multichannel
devices based on the valley properties \cite{valleyproperty_1,valley_device_1,valley_device_2,valley_device_3,valley_device_4,valley_device_5}.

Recognized as one type of valleytronic materials, 2-dimensional (2D)
group-IV monochalcogenids AB (A=Ge,Sn;B=Se,S) have received extensive
attentions \cite{AB_monogap,AB_monomemory,AB_monopolarization,AB_nanoletter,DS_3,DS_4,DS_5,gese1,gese2,gese3,gese4,gese5}.
Experimentally, group-IV monochalcogenids adopt an orthorhombic structure
($Pnma$ space group) as shown in Fig. 1(a), where one unit cell is constructed
by two puckered polar layers similar to that of black phosphorus~\cite{blackP_absorb1,blackP_absorb2,blackP_absorb3}.
So $Pnma$ monochalcogenids can be exfoliated to the monolayer
\cite{AB_monomemory,exfoliate_1}, and feature multiple valleys
\cite{MultiValley1}, distinctive optical selection rules \cite{OpticalSelection1,OpticalSelection2},
extraordinary spintronics \cite{spintronics}, large piezoelectric
\cite{piezoelectric} and ferroelectric effects \cite{ferroelectric_1,DV_2,DV_3}.

\begin{figure}
\includegraphics[scale=0.5]{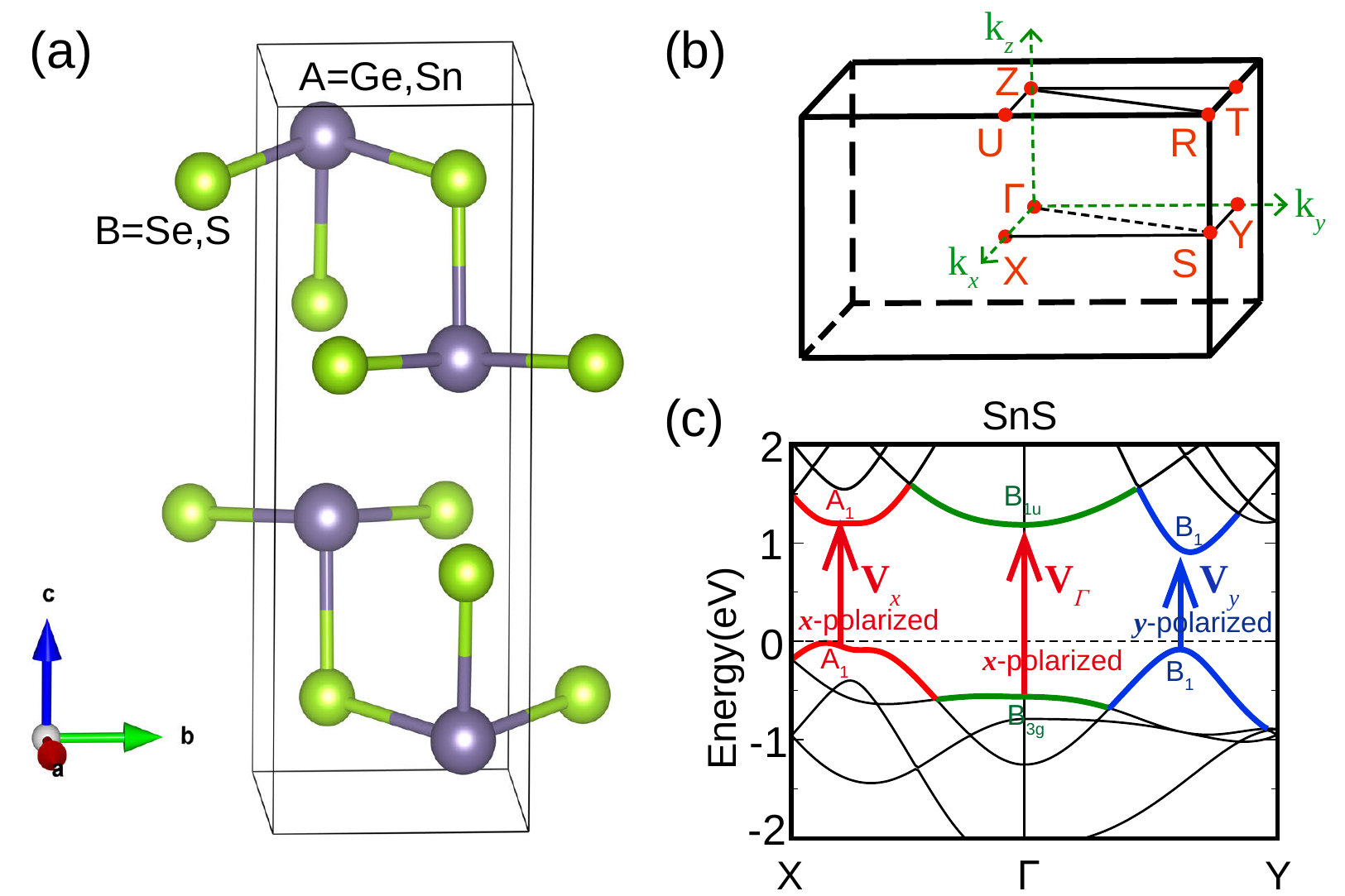}\protect\protect\protect\caption{(a) Crystal structure of AB compounds, which contains two puchered
polar layers per unit cell. (b) The first Brillouin zone (BZ) of AB
compounds, in which the high symmetry points are labeled. (c) Valley
definition and their optical absorption rules in AB compounds, where
the irrep for each valley are also marked by $A_{1}$, $B_{1u}$,
$B_{3g}$ and $B_{1}$.\label{fig:Crystal-structure}}
\end{figure}

While valleytronics are widely studied in the 2D materials, the valley physics and their polarization generating in 3-dimensional (3D) compounds is rarely achieved. In this paper, by means of the density functional theory (DFT) calculations and symmetry analysis, we demonstrate that three pairs of valleys exist at the Fermi level in the bulk of group-IV monochalcogenids, which feature different selection rules to the linearly-polarized light. The linear polarization-resolved infrared (IR) transmission spectra are measured on the GeSe crystal.
An obvious two-fold linear dichroism absorption between 8600 to 10000 $cm^{-1}$ is observed, corresponding to our calculations and optical absorption analyses very well. Such observation demonstrates that the valley polarization is successfully generated in the 3D group-IV monochalcogenids.
Furthermore, the valley position versus the in-plane strain is also studied in this work, which offers a useful method to optimize the valleytronic properties in AB compounds.
All these results provide us a new direction for the exploring of the valley materials, as well as a great guidance for the design of the photodetector and valleytonic devices.

\section{\label{sec:level2}DFT calculations and band analysis}
Our DFT calculations are
performed by Vienna \emph{ab initio} Simulation Package (VASP) within
the projector augmented wave scheme \cite{CAL_VASP}, where the generalized
gradient approximation (GGA) of the Perdew-Burke-Ernzerhof type for
the exchange-correlation potential \cite{DV_1,CAL_PBE} are adopted. The
cutoff energy for the wave function expansion is set to 520 eV. A
Monkhorst-Pack grid of 10$\times$10$\times$4 \emph{k}-meshes are
used for the self-consistent calculations\cite{CAL_KPOINTS}. Since
the spin-orbital coupling (SOC) has very weak influence on the band
structures for the existence of the inversion symmetry, we have excluded
SOC in all calculations, as well as our optical selection rule
analyses.

\begin{table}
\centering{}\protect\caption{Crystal parameters of AB family used in our calculations.}
\begin{tabular}{c|ccccc}
\hline
 & $a$(${\AA}$)  & $b$(${\AA}$)  & $c$(${\AA}$)  & A1  & B1\tabularnewline
\hline
\hline
GeSe  & 4.45  & 3.85  & 10.76  & (0.10,0.25,0.87)  & (0.49,0.25,0.15)\tabularnewline
\hline
GeS  & 4.43  & 3.65  & 10.43  & (0.12,0.25,0.87)  & (0.50,0.25,0.15)\tabularnewline
\hline
SnSe  & 4.46  & 4.19  & 11.58  & (0.12,0.25,0.87)  & (0.50,0.25,0.15)\tabularnewline
\hline
SnS  & 4.33  & 3.98  & 11.18  & (0.12,0.25,0.87)  & (0.48,0.25,0.15)\tabularnewline
\hline
\end{tabular}\label{tab:crystaldata}
\end{table}

Experimental crystal parameters shown in Table I are used in our calculations.
The calculated band structures of AB compounds are plotted in Fig.
2, which indicate that four AB compounds are all insulators, and
have very similar band dispersions and orbital characters around the
Fermi level. Therefore, we would like to take the bands of SnS as
a representative to demonstrate their orbital components and valley
definitions in the following. Firstly, the insulating gap for all compounds
are determined by the band dispersions in the $k_{z}=0$ plane of
the Brillouin zone (BZ), as shown in the Fig. 2. Especially, there
are three energy-adjacent minima (maxima) located on the conduction
(valence) band along the $X-\Gamma-Y$ directions, forming three pairs
of valleys, $V_{x}$ confined in $\Gamma-X$ direction, $V_{y}$ confined
in $\Gamma-Y$ direction, and $V_{\Gamma}$ located around $\Gamma$
point, respectively. The energy
difference between $E_{V_{x}}$ and $E_{V_{y}}$ is defined as $\Delta E=E_{V_{y}}-E_{V_{x}}$. Secondly, due to the symmetry requirement, $V_{x}$,
$V_{y}$, and $V_{\Gamma}$ consist of different orbits
and belong to different representations. As shown in Fig. 1(c) and Fig. 2, valleys $V_{x}$ colored by red and valleys $V_{y}$
colored by blue are composed of $p_{x}$ and $p_{y}$ orbitals respectively
due to the existence of $C_{2z}$ rotation perpendicular to the $xy$
plane, while valleys $V_{\Gamma}$ represented by the green bands, are mainly
composed of $p_{z}$ orbitals. The different band compositions would
lead to distinctive optical absorption rules, which are summarized
in Fig. 1(c), and will be analyzed in the next chapter.

\begin{figure}
\includegraphics[scale=0.45]{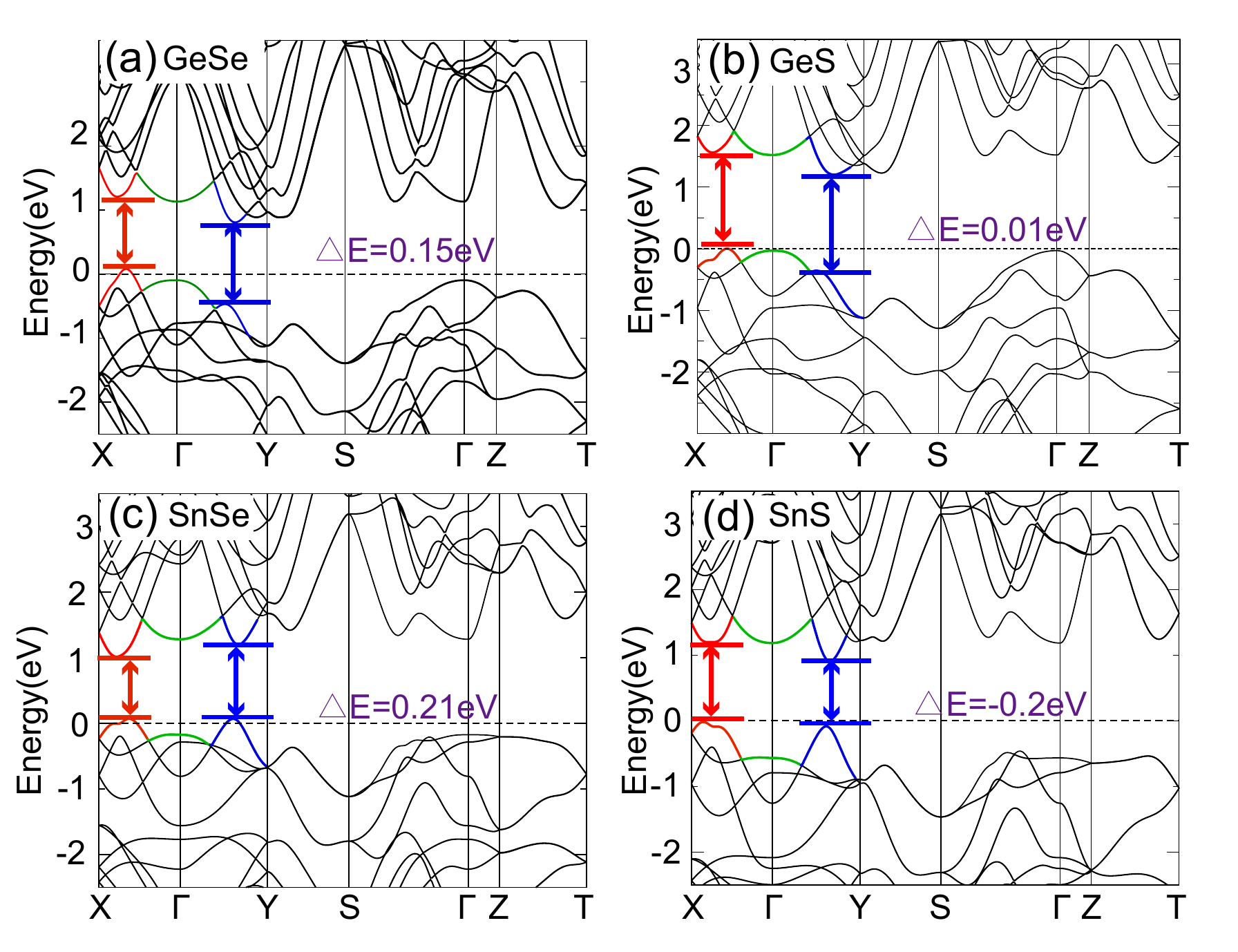}\protect\protect\protect\caption{ The calculated band structures of AB compounds, in which the energy
difference between $E_{V_{x}}$ and $E_{V_{y}}$ is defined by $\Delta E=E_{V_{y}}-E_{V_{x}}$.}

\label{fig:Bandstructures-for-AB}
\end{figure}

\section{\label{sec:level3}Optical selection rules}
As we all know, the interband transition
probability $P_{i}(\mathbf{k})$ is proportional to $\left|\left\langle c(\mathbf{k})\left|E_{i}\right|v(\mathbf{k})\right\rangle \right|^{2}\propto\left|\left\langle c(\mathbf{k})\left|\frac{\partial H}{\partial k_{i}}\right|v(\mathbf{k})\right\rangle \right|^{2}$\cite{valley_device_5},
where $E_{i}$ is the direction of the electric field of the polarized
light, $c(\mathbf{k})$ and $v(\mathbf{k})$ are the conduction and
valence band's wave functions at a given momentum $\mathbf{k}$,
respectively. The electron excitation from the valence band to the
conduction band could happen only when the integral of $\left\langle c(\mathbf{k})\left|\frac{\partial H}{\partial k_{i}}\right|v(\mathbf{k})\right\rangle $
is nonzero. It means that the product of $c(\mathbf{k})$, $\frac{\partial H}{\partial k_{i}}$
and $v(\mathbf{k})$ should be unchanged under any symmetry operations of
the system, \emph{i.e.} their irreducible representations' (irreps)
product must be identical ($A_{1g}$ in Table II and $A_{1}$ in Table
III).

Let us first address the selection rule of the valley located around
the $\Gamma$ point, $V_{\Gamma}$, which has never been studied before.
According to our DFT calculations, the composition of the highest
valence band and the lowest conduction band around the $\Gamma$ point
are both $p_{z}$ orbitals, and their irreps are $B_{3g}$ and $B_{1u}$
with respect to the $D_{2h}$ little group, as shown in Table II.
It means that only the $B_{2u}$ irrep satisfies $B_{1u}\otimes B_{2u}\otimes B_{3g}=A_{1g}$,
\emph{i.e.}, only the $x$-polarized light could excite the electrons
transition between the $V_{\Gamma}$ valleys. For the $V_{x}$ and
$V_{y}$ valleys, their little group are both $C_{2v}$. Two $V_{x}$
valleys (red bands in Fig. 1 and Fig. 2) located along $\Gamma-X$
direction both belong to $A_{1}$ irrep, which decides that only the
$x$-polarized light with $A_{1}$ irrep could excite the $V_{x}$
polarization, yielding to the requirement $A_{1}\otimes A_{1}\otimes A_{1}=A_{1}$.
On the other hand, two $V_{y}$ valleys located along $\Gamma-Y$
direction both have $B_{1}$ irreps. Thus, only the $y$-polarized
light could excite the $V_{y}$ polarization due to $B_{1}\otimes A_{1}\otimes B_{1}=A_{1}$.
The optical absorption rules of $V_{x}$, $V_{y}$ and $V_{\Gamma}$
are summarized in Fig. 1(c). Using these features, we can distinguish
the direct energy gaps of $V_{x}$ and $V_{y}$ valleys. We note that the energy gap between the $V_{\Gamma}$ valleys,
\emph{i.e.} $E_{V_{\Gamma}}$, is always larger than that between $V_{x}$
valleys ($E_{V_{x}}$) based on our calculations. So we can
identify $E_{V_{x}}$ by observing the absorption edge of the $x$-polarized
light. Furthermore, two kinds of valley polarization can be generated,
since $V_{x}$ and $V_{y}$ can be pumped separately by applying the $x/y$-polarized light.

\begin{figure}
\includegraphics[scale=0.4]{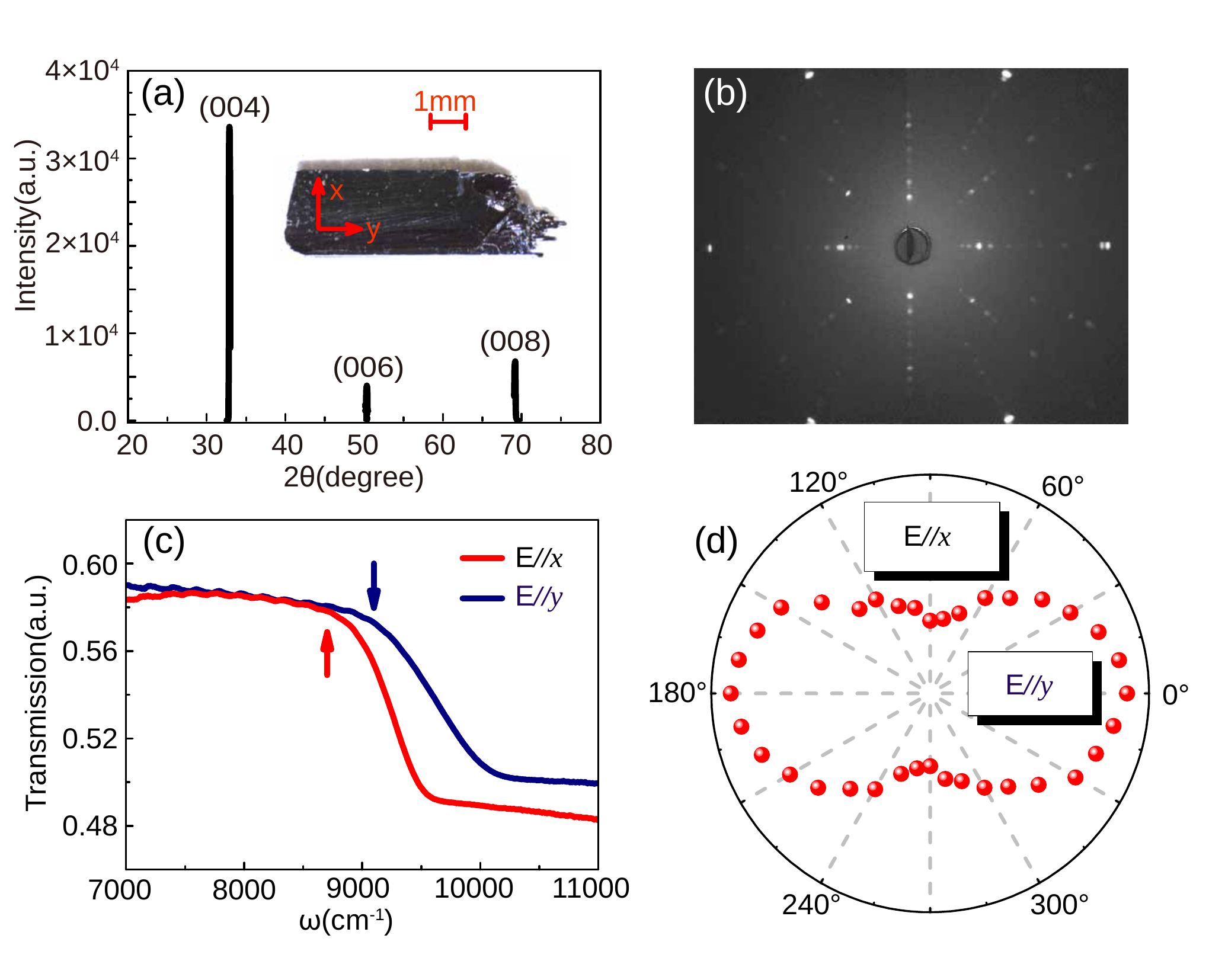}\protect\protect\protect\caption{\label{fig:GESEcrystal}(a) Single crystal XRD pattern of $xy$-plane
in GeSe. Inset: Optical micrograph of GeSe crystal. (b) Lane diffraction pattern of GeSe crystal. (c) Linear polarization-resolved IR
transmission spectra when light is polarized along $x$ and $y$ directions.
(d) Energy gaps measured from $0^{\circ}$ to $360^{\circ}$ on GeSe
and plotted in a polar coordinate.}
\end{figure}

\section{\label{sec:level4}Valley polarization generation}
Guiding by the above analysis,
we have grown GeSe and GeS compounds, and measured the linearly-polarized
IR transmission spectra to verify the the polarization-dependent
absorption (linear dichroism), as well as the generation of the valley
polarization in the 3D group-IV monochalcogenids. Here we take GeSe
as an example to illustrate such linear dichroism and the valley polarization
generation.

Our GeSe crystals are grown by the chemical vapor transport(CVT)
technique with iodine as the transporting agent \cite{CVT_1,CVT_2}.
High purity germanium ($>99.999\%$) and selenium ($>99.99\%$) powders
with 1:1 composition are thoroughly mixed and sealed together with
iodine (4 mg/liter) in a silica ampule under pressure of $10^{-3}$
Pa. The sealed ampule is placed into a two stage horizontal furnace.
The hot end with the mixture is slowly heated to 820 K at the rate
of 35 K/h and the cold end to 770 K at the rate of 32 K/h. The slow
heating rate is necessary to avoid any explosion due to the strong
exothermic reaction between the elements. After 120 hours of growth,
the furnace is cooled down naturally. Single crystals of GeSe with
the size of $\sim1.5\times5\times0.5mm^{3}$ are obtained after breaking
the quartz ampoule, as shown in the inset of Fig.\ref{fig:GESEcrystal}(a).

\begin{table}[tp]
\centering{}\protect\protect\caption{Irreducible representations of AB compounds at $\Gamma$ point with
$D_{2h}$ little group.}

\begin{tabular}{c|ccccccccc}
\hline
 & $E$  & $C_{2z}$  & $C_{2y}$  & $C_{2x}$  & $I$  & $M_{z}$  & $M_{y}$  & $M_{x}$  & function\tabularnewline
\hline
\hline
$A_{1g}$  & 1  & 1  & 1  & 1  & 1  & 1  & 1  & 1  & 1\tabularnewline
\hline
$B_{3g}$  & 1  & -1  & 1  & -1  & 1  & -1  & 1  & -1  & $y$\tabularnewline
\hline
$B_{1u}$  & 1  & 1  & -1  & -1  & -1  & -1  & 1  & 1  & $z$\tabularnewline
\hline
$B_{2u}$  & 1  & -1  & -1  & 1  & -1  & 1  & 1  & -1  & $x$\tabularnewline
\hline
\end{tabular}\label{tab:gamma}
\end{table}

\begin{table}
\centering{}\protect\protect\caption{Irreps of the $C_{2v}$ little group for the $V_{x}$ and $V_{y}$
valleys in AB compounds. $C_{2}$ is the two-fold rotation around
the $x/y$-axis; $M_{y/x}$ is the mirror reflection with respect
to the plane perpendicular to the $y/x$-axis; and $M_{z}$ means
the mirror reflection with respect to the plane perpendicular to the
$z$-axis. The sixth and seventh column are the functions for $V_{x}$
and $V_{y}$, respectively.}

\centering{}
\begin{tabular}{c|cccccc}
\hline
 & $E$  & $C_{2}$  & $M_{y/x}$  & $M_{z}$  & function\_$V_{x}$  & function\_$V_{y}$ \tabularnewline
\hline
\hline
$A_{1}$  & 1  & 1  & 1  & 1  & 1  & 1~\tabularnewline
\hline
$B_{1}$  & 1  & -1  & -1  & 1  & $x$ & $y$~\tabularnewline
\hline
$B_{2}$  & 1  & -1  & 1  & -1  & $z$ & $z$~\tabularnewline
\hline
$A_{2}$  & 1  & 1  & -1  & -1  & $y$ & $x$~\tabularnewline
\hline
\end{tabular}\label{tab:Irreducible-representations-of}
\end{table}

The as-grown crystals are characterized by the scanning electron
microscopy (SEM) with energy dispersive spectrometer (EDS), single
crystal X-ray diffraction (XRD) and Laue back reflection measurements.
Stoichiometric GeSe composition is confirmed by the EDS analysis.
The flat surface is identified to be (0 0 1) plane by the single crystal
XRD measurements, as displayed in Fig.\ref{fig:GESEcrystal}(a). It
indicates that the flat surface is perpendicular to the crystalline
$c$-axis (defined as $z$-direction), therefore it is the \emph{ab}-plane
in the structure \cite{XRD,XRD_2}. The $a$- and $b$-axis ($x$-
and $y$-direction) in this surface are determined by Laue back reflection
camera system \cite{Lauebackreflectioncamera}, which is shown in Fig.\ref{fig:GESEcrystal}(b).

The linear polarization-resolved IR transmission spectra
of GeSe crystals are measured at room temperature using a Bruker
IFS 80v spectrometer. The incident beam is along the $z$ direction
and the linearly-polarized electric field could be turned within the
$xy$-plane. Fig.\ref{fig:GESEcrystal}(c) shows the transmission
spectra with electric field along the $x$- and $y$-direction, respectively.
The sharp drop signals the onset of the absorption due to
that the photon energy exceeds the band gap, which can be used to
identify the value of the direct valley energy gap, $E_{V_{x}}$ and
$E_{V_{y}}$. The measurements demonstrate a clear linear dichroism
between 8600 to 10000 $cm^{-1}$. For the $x$-polarized light, the
onset of absorption is about 8600 $cm^{-1}$ (or 1.07 eV), corresponding
to our calculated $E_{V_{x}}=1.12$ eV. By contrast, the onset absorption
of the $y$-polarized light is enhanced to about 9100 $cm^{-1}$ (or
1.13 eV), corresponding to our calculated $E_{V_{y}}=1.27$ eV roughly.
The observed difference of $E_{V_{x}}$ and $E_{V_{y}}$ is 0.06 eV,
a little smaller than our calculated result 0.15eV,
but declaring the existence of the linear dichroism absorption enough.
To further resolve the linear dichroism, we measured the polarization-dependent
IR transmission spectra of the same sample with the normal light incidence
and the polarization angles relative to $x$-direction from $0^{\circ}$
to $360^{\circ}$ in steps of $10^{\circ}$. Fig.\ref{fig:GESEcrystal}(d)
shows the measured band gaps in a polar coordinate. A
two-fold anisotropy along the $x$ and $y$ directions could be clearly
seen in the plot, reflecting the electronic anisotropy along the $\Gamma-X$
and $\Gamma-Y$ directions.

The observed linear dichroism in GeSe perfectly confirm our DFT
calculations and the optical absorption analyses. More importantly,
it also demonstrates that the valley polarization has been successfully
generated by the linearly-polarized light in group-IV monochalcogenids, which is realized in a 3D
system for the first time. Such valley polarization
can be detected by the converse process, or by using the nonlinear
transverse valley conductivity. Therefore our results provide new functionalities in
optical switches and valleytonics devices of the 3D group-IV monochalcogenids.

\begin{figure}
\includegraphics[scale=0.4]{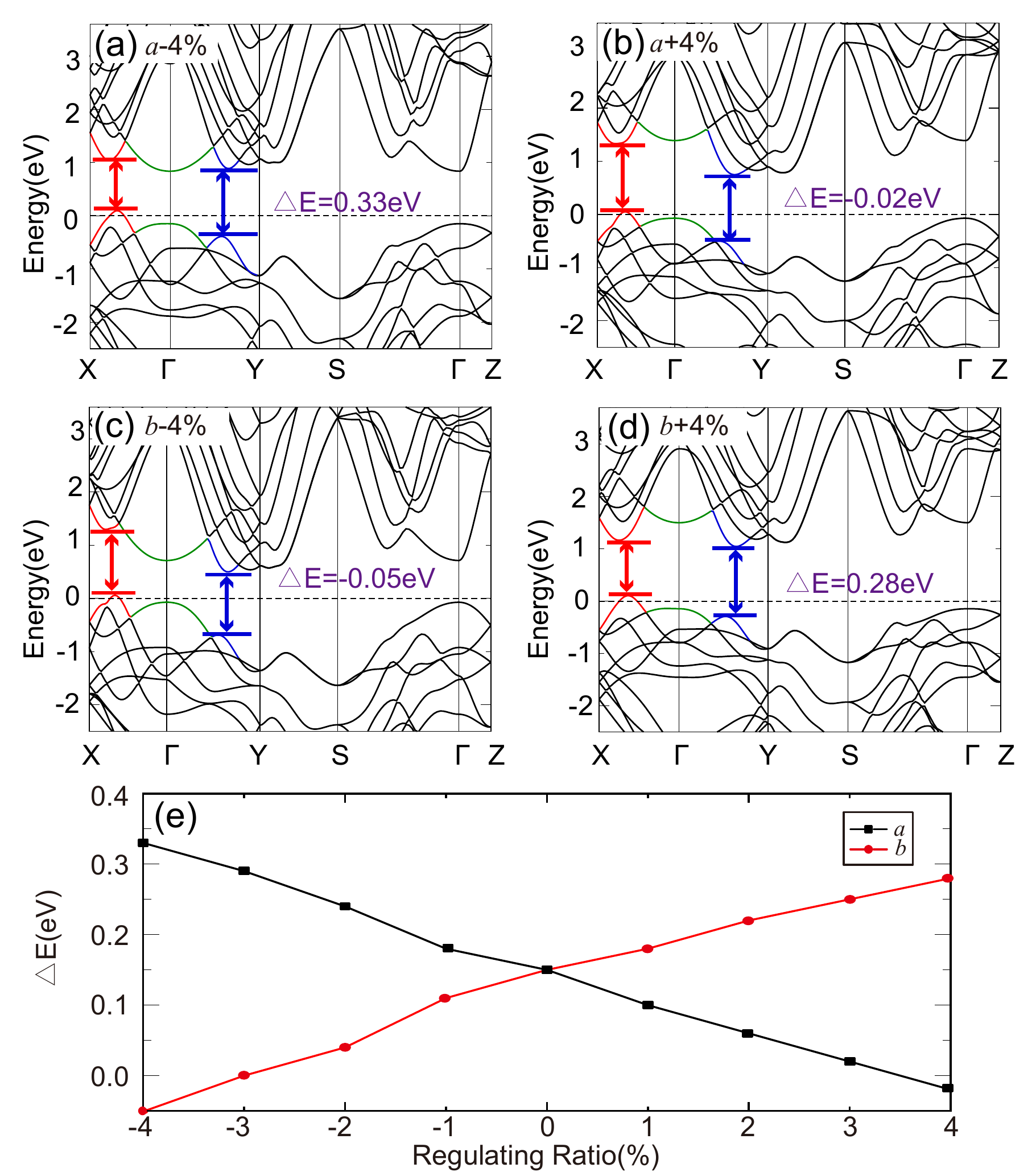}\protect\protect\protect\caption{\label{fig:abc}Band structrues and energy gap difference tendency
for GeSe after modulating crystal constants. (a-d) Band structures of GeSe
by reducing or enlarging $4\%$ of crystal constants $a$ and $b$.
(e) The evolution of the energy gap difference $\Delta E$ as function of the lattice constants $a$ (black) and $b$ (red), respectively.}
\end{figure}

\section{\label{sec:level5}Valley modulation}
To enhance the performance of the optical
switches and the valley polarization, one expects that the energy difference
$\Delta E$ should be as large as possible, and the $V_{\Gamma}$ valleys
should be away from the Fermi level as far as they can. Based on
our calculations, SnS satisfies these requirements mostly, which is
the best 3D valley material among group-IV monochalcogenides. On the
other hand, we can modify the valley position by modulating the external
conditions, such as strain \cite{ModulationStrain1,ModulationStrain2},
gating field \cite{ModulationGatingfield1,ModulationGatingfield2}
\emph{et al}. Since $E_{V_{x}}$ and
$E_{V_{y}}$ are decided by the energy positions of the $p_{x}$ and
$p_{y}$ orbitals. One would naturally expect that the in-plane strains
are one of the effective methods to modulate $\Delta E$. In Fig.4 (a-d),
we have plotted the band structures of GeSe by reducing or enlarging
$4\%$ of $a$-axis and $b$-axis respectively, which confirms that
reducing a-axis or enlarging $b$-axis can both support a larger linear
dichroism energy window, \emph{i.e.} a larger $\Delta E$. The $\Delta E$
evolution of the $a$-axis ($b$-axis) stain is summarized in Fig.
4(e), from which we can see that either reducing constant $a$ or
enlarging constant $b$ could increase the $\Delta E$ monotonically.
However, for the case of reducing $a$-axis, those $V_{\Gamma}$ valleys
become very close to the Fermi level as shown in Fig. 4(a), which
is disadvantageous to the $V_{x}$ polarization with the pure $x$-direction
momentum. Therefore, we conclude that enlarging $b$-axis is the ideal
way to get the best valleytronic properties in GeSe.

\section{\label{sec:level6}Conclusions}
Combining the DFT calculations and the linear polarization-resolved
IR transmission measurement together, we have proposed a family
of 3D valleytronic materials AB (A=Ge, Sn; B=S,Se), in which a distinctive
two-fold anisotropic absorption of the linearly-polarized light is
observed, consistent with the optical absorption rule analyses very well. The observed linear
dichroism perfectly confirmed that valley polarization is successfully
generated by the linearly-polarized light in a 3D system. Such valley polarization can be read by using the nonlinear
transverse valley conductivity. Furthermore, valley modulation
by the in-plane strain is also studied in our work, which points out
a useful method to optimize the vallaytronic properties in AB compounds.
All these results open a new direction for the exploration of the valley materials,
which also provide a good platform of optical switches and valleytonic devices.

Thank Zhida Song for useful discussion. This work is supported by
the National Thousand-Young-Talents Program and the NSFC. S.-J. Z and N.-L. W are supported by supported by the National Science Foundation of China (No. 11327806), the National Key Research and Development Program of China (No.2016YFA0300902).

\bibliographystyle{unsrt}

\begin{thebibliography}{48}

\bibitem{valleypolarization_1}
JL~Garcia-Pomar, A~Cortijo, and M~Nieto-Vesperinas.
\newblock Fully valley-polarized electron beams in graphene.
\newblock {\em Physical review letters}, 100(23):236801, 2008.

\bibitem{valleyproperty_1}
Di~Xiao, Gui-Bin Liu, Wanxiang Feng, Xiaodong Xu, and Wang Yao.
\newblock Coupled spin and valley physics in monolayers of MoS2 and other
  group-VI dichalcogenides.
\newblock {\em Physical Review Letters}, 108(19):196802, 2012.

\bibitem{Valleytronics_1}
Kamran Behnia.
\newblock Condensed-matter physics: polarized light boosts valleytronics.
\newblock {\em Nature nanotechnology}, 7(8):488--489, 2012.

\bibitem{Valleytronics_2}
Christoph~E Nebel.
\newblock Valleytronics: Electrons dance in diamond.
\newblock {\em Nature materials}, 12(8):690--691, 2013.

\bibitem{Opticalabsorption_1}
Hualing Zeng, Junfeng Dai, Wang Yao, Di~Xiao, and Xiaodong Cui.
\newblock Valley polarization in MoS2 monolayers by optical pumping.
\newblock {\em Nature nanotechnology}, 7(8):490--493, 2012.

\bibitem{Opticalabsorption_2}
Kin~Fai Mak, Keliang He, Jie Shan, and Tony~F Heinz.
\newblock Control of valley polarization in monolayer MoS2 by optical helicity.
\newblock {\em Nature nanotechnology}, 7(8):494--498, 2012.

\bibitem{Opticalabsorption_3}
Kamran Behnia.
\newblock Condensed-matter physics: polarized light boosts valleytronics.
\newblock {\em Nature nanotechnology}, 7(8):488--489, 2012.

\bibitem{Opticalabsorption_4}
Ajit Srivastava, Meinrad Sidler, Adrien~V Allain, Dominik~S Lembke, Andras Kis,
  and A~Imamo{\u{g}}lu.
\newblock Valley zeeman effect in elementary optical excitations of monolayer
  WSe2.
\newblock {\em Nature Physics}, 2015.

\bibitem{Opticalabsorption_5}
Yilei Li.
\newblock Valley splitting and polarization by zeeman effect in monolayer
  MoSe2.
\newblock In {\em Probing the Response of Two-Dimensional Crystals by Optical
  Spectroscopy}, pages 55--64. Springer, 2016.

\bibitem{tuningproperties_1}
Honglai Li, Xidong Duan, Xueping Wu, Xiujuan Zhuang, Hong Zhou, Qinglin Zhang,
  Xiaoli Zhu, Wei Hu, Pinyun Ren, Pengfei Guo, et~al.
\newblock Growth of alloy MoS2xSe2(1-x) nanosheets with fully tunable
  chemical compositions and optical properties.
\newblock {\em Journal of the American Chemical Society}, 136(10):3756--3759,
  2014.

\bibitem{tuningproperties_2}
Qing~Hua Wang, Kourosh Kalantar-Zadeh, Andras Kis, Jonathan~N Coleman, and
  Michael~S Strano.
\newblock Electronics and optoelectronics of two-dimensional transition metal
  dichalcogenides.
\newblock {\em Nature nanotechnology}, 7(11):699--712, 2012.

\bibitem{valley_device_1}
Qing~Hua Wang, Kourosh Kalantar-Zadeh, Andras Kis, Jonathan~N Coleman, and
  Michael~S Strano.
\newblock Electronics and optoelectronics of two-dimensional transition metal
  dichalcogenides.
\newblock {\em Nature nanotechnology}, 7(11):699--712, 2012.

\bibitem{valley_device_2}
Priya Johari and Vivek~B Shenoy.
\newblock Tuning the electronic properties of semiconducting transition metal
  dichalcogenides by applying mechanical strains.
\newblock {\em ACS nano}, 6(6):5449--5456, 2012.

\bibitem{valley_device_3}
Manish Chhowalla, Hyeon~Suk Shin, Goki Eda, Lain-Jong Li, Kian~Ping Loh, and
  Hua Zhang.
\newblock The chemistry of two-dimensional layered transition metal
  dichalcogenide nanosheets.
\newblock {\em Nature chemistry}, 5(4):263--275, 2013.

\bibitem{valley_device_4}
Hualing Zeng, Junfeng Dai, Wang Yao, Di~Xiao, and Xiaodong Cui.
\newblock Valley polarization in MoS2 monolayers by optical pumping.
\newblock {\em Nature nanotechnology}, 7(8):490--493, 2012.

\bibitem{valley_device_5}
Di~Xiao, Gui-Bin Liu, Wanxiang Feng, Xiaodong Xu, and Wang Yao.
\newblock Coupled spin and valley physics in monolayers of MoS2 and other
  group-VI dichalcogenides.
\newblock {\em Physical Review Letters}, 108(19):196802, 2012.

\bibitem{AB_monogap}
L{\'\i}dia~C Gomes and A~Carvalho.
\newblock Phosphorene analogues: Isoelectronic two-dimensional group-IV
  monochalcogenides with orthorhombic structure.
\newblock {\em Physical Review B}, 92(8):085406, 2015.

\bibitem{AB_monomemory}
Paul~Z Hanakata, Alexandra Carvalho, David~K Campbell, and Harold~S Park.
\newblock Memory effects in monolayer group-IV monochalcogenides.
\newblock {\em Stress (GPa)}, 1(1.5):2, 2016.

\bibitem{AB_monopolarization}
Paul~Z Hanakata, Alexandra Carvalho, David~K Campbell, and Harold~S Park.
\newblock Polarization and valley switching in monolayer group-IV
  monochalcogenides.
\newblock {\em Physical Review B}, 94(3):035304, 2016.

\bibitem{AB_nanoletter}
Guangsha Shi and Emmanouil Kioupakis.
\newblock Anisotropic spin transport and strong visible-light absorbance in
  few-layer SnSe and GeSe.
\newblock {\em Nano letters}, 15(10):6926--6931, 2015.

\bibitem{DS_3}
Lijun Zhang and D.~J. Singh.
\newblock Electronic structure and thermoelectric properties of layered
  ${\text{PbSe-WSe}}_{2}$ materials.
\newblock {\em Phys. Rev. B}, 80:075117, Aug 2009.

\bibitem{DS_4}
David Parker and David~J. Singh.
\newblock High temperature thermoelectric properties of rock-salt structure
  PbS.
\newblock {\em Solid State Communications}, 182:34 -- 37, 2014.

\bibitem{DS_5}
David Parker and David~J Singh.
\newblock First principles investigations of the thermoelectric behavior of tin sulfide.
\newblock {\em Journal of Applied Physics}, 108(8):083712, 2010.

\bibitem{gese1}
Rangel, Tonatiuh and Fregoso, Benjamin M. and Mendoza, Bernardo S. and Morimoto, Takahiro and Moore, Joel E. and Neaton, Jeffrey B..
\newblock Large Bulk Photovoltaic Effect and Spontaneous Polarization of Single-Layer Monochalcogenides.
\newblock{\em Phys. Rev. Lett.}, 119(6):067402, 2017

\bibitem{gese2}
Bertrand, Simon and Garate, Ion and C\^ot\'e, Ren\'e.
\newblock Light-induced valley polarization in interacting and nonlinear Weyl semimetals.
\newblock{\em Phys. Rev. B}, 96(7):075126, 2017

\bibitem{gese3}
{Bertrand}, S. and {Garate}, I. and {C{\^o}t{\'e}}, R..
\newblock Optical absorption in interacting and nonlinear Weyl semimetals.
\newblock{\em arXiv}, 1704.08939, 2017

\bibitem{gese4}
{Pletikosi{\'c}}, I. and {von Rohr}, F. and {Pervan}, P. and {Das}, P.~K. and {Vobornik}, I. and {Cava}, R.~J. and {Valla}, T..
\newblock Band structure of a IV-VI black phosphorus analogue, the thermoelectric SnSe.
\newblock{\em arXiv}, 1707.04289,2017

\bibitem{gese5}
von Rohr, Fabian O. and Ji, Huiwen and Cevallos, F. Alexandre and Gao, Tong and Ong, N. Phuan and Cava, Robert J..
\newblock High-Pressure Synthesis and Characterization of ¦Â-GeSe¡ªA Six-Membered-Ring Semiconductor in an Uncommon Boat Conformation.
\newblock{\em Journal of the American Chemical Society}, 139(7):2771-2777,2017






\bibitem{blackP_absorb1}
Likai Li, Yijun Yu, Guo~Jun Ye, Qingqin Ge, Xuedong Ou, Hua Wu, Donglai Feng,
  Xian~Hui Chen, and Yuanbo Zhang.
\newblock Black phosphorus field-effect transistors.
\newblock {\em Nature nanotechnology}, 9(5):372--377, 2014.

\bibitem{blackP_absorb2}
Fengnian Xia, Han Wang, and Yichen Jia.
\newblock Rediscovering black phosphorus as an anisotropic layered material for
  optoelectronics and electronics.
\newblock {\em Nature communications}, 5, 2014.

\bibitem{blackP_absorb3}
Jingsi Qiao, Xianghua Kong, Zhi-Xin Hu, Feng Yang, and Wei Ji.
\newblock High-mobility transport anisotropy and linear dichroism in few-layer
  black phosphorus.
\newblock {\em Nature communications}, 5, 2014.

\bibitem{exfoliate_1}
Jack~R Brent, David~J Lewis, Tommy Lorenz, Edward~A Lewis, Nicky Savjani,
  Sarah~J Haigh, Gotthard Seifert, Brian Derby, and Paul O¡¯Brien.
\newblock Tin (ii) sulfide(SnS) nanosheets by liquid-phase exfoliation of
  herzenbergite: IV-VI main group two-dimensional atomic crystals.
\newblock {\em Journal of the American Chemical Society}, 137(39):12689--12696,
  2015.

\bibitem{MultiValley1}
S.~Gwo, K.-J. Chao, C.~K. Shih, K.~Sadra, and B.~G. Streetman.
\newblock Direct mapping of electronic structure across
  ${\mathrm{Al}}_{0.3}$${\mathrm{Ga}}_{0.7}$As/GaAs heterojunctions: Band
  offsets, asymmetrical transition widths, and multiple-valley band structures.
\newblock {\em Phys. Rev. Lett.}, 71:1883--1886, Sep 1993.

\bibitem{OpticalSelection1}
Yu-xi Liu, J.~Q. You, L.~F. Wei, C.~P. Sun, and Franco Nori.
\newblock Optical selection rules and phase-dependent adiabatic state control
  in a superconducting quantum circuit.
\newblock {\em Phys. Rev. Lett.}, 95:087001, Aug 2005.

\bibitem{OpticalSelection2}
P.~Voisin, G.~Bastard, and M.~Voos.
\newblock Optical selection rules in superlattices in the envelope-function
  approximation.
\newblock {\em Phys. Rev. B}, 29:935--941, Jan 1984.

\bibitem{spintronics}
David~D Awschalom and Michael~E Flatt{\'e}.
\newblock Challenges for semiconductor spintronics.
\newblock {\em Nature Physics}, 3(3):153--159, 2007.

\bibitem{piezoelectric}
Wanling Pan and Farrokh Ayazi.
\newblock Thin-film piezoelectric-on-substrate resonators with Q enhancement
  and TCF reduction.
\newblock In {\em Micro Electro Mechanical Systems (MEMS), 2010 IEEE 23rd
  International Conference on}, pages 727--730. IEEE, 2010.

\bibitem{ferroelectric_1}
Guo-Xing Miao and Jagadeesh~S Moodera.
\newblock Spin manipulation with magnetic semiconductor barriers.
\newblock {\em Physical Chemistry Chemical Physics}, 17(2):751--761, 2015.

\bibitem{DV_2}
R.~D. King-Smith and David Vanderbilt.
\newblock Theory of polarization of crystalline solids.
\newblock {\em Phys. Rev. B}, 47:1651--1654, Jan 1993.

\bibitem{DV_3}
Fabio Bernardini, Vincenzo Fiorentini, and David Vanderbilt.
\newblock Polarization-based calculation of the dielectric tensor of polar
  crystals.
\newblock {\em Phys. Rev. Lett.}, 79:3958--3961, Nov 1997.

\bibitem{CAL_VASP}
G~Kresse, J. Furthm{\"u}ller.
\newblock Software VASP, Vienna, 1999.


\bibitem{DV_1}
David Vanderbilt.
\newblock Soft self-consistent pseudopotentials in a generalized eigenvalue
  formalism.
\newblock {\em Phys. Rev. B}, 41:7892--7895, Apr 1990.

\bibitem{CAL_PBE}
Matthias Ernzerhof and Gustavo~E Scuseria.
\newblock Assessment of the Perdew--Burke--Ernzerhof exchange-correlation
  functional.
\newblock {\em The Journal of chemical physics}, 110(11):5029--5036, 1999.

\bibitem{CAL_KPOINTS}
Hendrik~J Monkhorst and James~D Pack.
\newblock Special points for brillouin-zone integrations.
\newblock {\em Physical review B}, 13(12):5188, 1976.

\bibitem{CVT_1}
Heribert Wiedemeier, Eugene~A Irene, and Asim~K Chaudhuri.
\newblock Crystal growth by vapor transport of GeSe, GeSe2, and GeTe and
  transport mechanism and morphology of GeTe.
\newblock {\em Journal of Crystal Growth}, 13:393 -- 396, 1972.

\bibitem{CVT_2}
Atsushi Okazaki.
\newblock The crystal structure of germanium selenide GeSe.
\newblock {\em Journal of the Physical Society of Japan}, 13(10):1151--1155,
  1958.

\bibitem{XRD}
Ajay Agarwal, P.~D. Patel, and D.~Lakshminarayana.
\newblock Single crystal growth of layered tin monoselenide semiconductor using
  a direct vapour transport technique.
\newblock {\em Journal of Crystal Growth}, 142(142):344--348, 1994.

\bibitem{XRD_2}
R.~Nitsche, H.~U. Blsterli, and M.~Lichtensteiger.
\newblock Crystal growth by chemical transport reactions¡ªi : Binary, ternary,
  and mixed-crystal chalcogenides.
\newblock {\em Journal of Physics Chemistry of Solids}, 21(s 3¨C4):199¨C205,
  1961.

\bibitem{Lauebackreflectioncamera}
HC~Hsueh, H~Vass, SJ~Clark, GJ~Ackland, and J~Crain.
\newblock High-pressure effects in the layered semiconductor germanium
  selenide.
\newblock {\em Physical Review B}, 51(23):16750, 1995.

\bibitem{ModulationStrain1}
E.~D. Minot, Yuval Yaish, Vera Sazonova, Ji-Yong Park, Markus Brink, and
  Paul~L. McEuen.
\newblock Tuning carbon nanotube band gaps with strain.
\newblock {\em Phys. Rev. Lett.}, 90:156401, Apr 2003.

\bibitem{ModulationStrain2}
Liangzhi Kou, Thomas Frauenheim, and Changfeng Chen.
\newblock Nanoscale multilayer transition-metal dichalcogenide
  heterostructures: Band gap modulation by interfacial strain and spontaneous
  polarization.
\newblock {\em The Journal of Physical Chemistry Letters}, 4(10):1730--1736,
  2013.
\newblock PMID: 26282986.

\bibitem{ModulationGatingfield1}
A.~V. Malyshev.
\newblock DNA double helices for single molecule electronics.
\newblock {\em Phys. Rev. Lett.}, 98:096801, Feb 2007.

\bibitem{ModulationGatingfield2}
S.~Caprara, F.~Peronaci, and M.~Grilli.
\newblock Intrinsic instability of electronic interfaces with strong rashba
  coupling.
\newblock {\em Phys. Rev. Lett.}, 109:196401, Nov 2012.

\end{thebibliography}

\end{document}